\documentclass[pre,aps,epsfig,twocolumn,showpacs]{revtex4}
\usepackage{graphicx}
\begin{document}

\title{ Stability of vacuum in coupled directed percolation processes}
\author{Sungchul Kwon}
\altaffiliation{also at School of Computational Sciences, KIAS}
\altaffiliation{Present address: Insititut f\"{u}r Festok\"{o}rperforshung, 
Forschungszentrum J\"{u}lich, 52425 J\"{u}lich, Germany}
\author{Hyunggyu Park}
\affiliation{School of Physics, Korea Institute for Advanced Study,
         Seoul 130-722, Korea}
\date{\today}

\begin{abstract}
We study the absorbing phase transitions in coupled directed percolation (DP)
processes with $N$-species particles in one dimension. 
The interspecies coupling is linear, bidirectional, and excitatory.  We find that the presence of a
spontaneous annihilation process $A\rightarrow 0$ 
is essential in stabilizing the absorbing phase (vacuum).  
In the coupled contact processes, the vacuum is stable and  the system exhibits DP type 
transitions, regardless of the coupling strength, for all $N$. However, in the coupled 
branching annihilation random walks with one offspring (BAW),
where particle annihilations occur only through binary diffusion processes
$A+A\rightarrow 0$, the vacuum becomes unstable with respect to an arbitrarily small branching
rate in a sufficiently strong coupling regime for $N\ge 3$.
The $N=2$  BAW exhibits the DP type transition for any coupling strength, but the inclusion of 
interspecies hard core (HC) interaction makes the vacuum unstable again and the system is
always active in a strong coupling regime. Critical behavior near the zero
branching point is characterized by the scaling exponents, 
$\beta=\nu_{\bot}=1/2$ and $\nu_{||}=1$, regardless of the presence of  HC
interaction. We also discuss the effects of the asymmetric coupling.
\end{abstract}

\maketitle
\section{Introduction}
Nonequilibrium systems with absorbing (trapped) states have been extensively studied 
in recent years, because of their wide applicability to various phenomena in natural science 
as well as in social and economical science  \cite{Marro,Haye}.  The absorbing transition between 
an active phase into an absorbing phase can be regarded as one of the simplest and natural 
extensions of the well-established equilibrium phase transition to nonequilibrium systems.

Critical behaviors near an absorbing transition are categorized into a few universality classes
characterized by the symmetry between absorbing states and/or the conservation in 
dynamics \cite{GKT,PP,Hinrichsen97}. The most prominent universality class is the directed 
percolation (DP) class \cite{JG}, which involves typically a single absorbing state without any 
conservation in dynamics. Most of absorbing-type nonequilibrium models belong to the DP class.
Only a small number of models form different classes such as the directed Ising (DI) class involving 
two equivalent classes of absorbing states \cite{GKT,PP,Meny} and the parity-conserving (PC) class 
with mod (2) conservation of the total particle number \cite{Taka}.
These two classes coincide in one dimension by identifying a domain wall in the DI-type models 
as a particle in the PC-type models. In higher dimensions, both types of models are always active
(no absorbing phase) except at a trivial point (annihilation fixed point) and 
their critical behavior is described by the mean field theory.  None of the models studied so far 
with higher symmetries than the Ising symmetry (for example, Potts symmetry)
stabilizes an absorbing phase even in one dimension.

Recently, a variety of coupled systems have been investigated extensively. 
Janssen \cite{Jans} studied coupled DP processes with bilinear and bidirectional interspecies
couplings in the framework of bosonic field theory, where no other critical phenomena 
were found than the DP.  T\"auber {\em et al} \cite{THH} studied linearly and unidirectionally coupled 
DP processes, where a series of new multicritical phenomena was observed. 
Coupled PC (DI) processes have been also studied \cite{CT}, where the absorbing phase become
unstable with respect to an arbitrarily small branching rate even in one dimension. 
More interestingly, the critical behavior near the annihilation fixed point depends on details of 
particle dynamics such as the presence of interspecies
hard core (HC) interaction and the branching method \cite{KLP,Odor1}. 

Stochastic models for linearly and bidirectionally coupled DP processes have been also studied
through two-species branching annihilating random walks with one offspring (BAW$_1$) in one dimension 
\cite{KP1,Odor2}.  It was found that the HC interaction is crucial in a strong coupling
regime, where the absorbing phase (vacuum) becomes unstable and the system is always active
except at the annihilation fixed point of zero branching rate. Critical behavior near the annihilation
fixed point was conjectured via an analytic argument and confirmed by numerical simulations
\cite{KP1}.

In this paper, we studied the $N$-species BAW$_1$ models and also the $N$-species contact processes (CP)
with and without interspecies HC interaction.  The single-species BAW$_1$ and CP both belong to the DP universality
class, but their upper critical dimensions are different ($d_{\rm uc}=2$ and 4) \cite{Taka,CT,recent}. 
The essential difference
is the absence of spontaneous annihilation process $A\rightarrow \emptyset$ in the BAW$_1$, where
particle annihilations occur only in pairs through binary diffusion processes $A+A\rightarrow \emptyset$.  
This annihilation random walk nature of the BAW$_1$ makes the vacuum unstable for $d\ge 2$.
In the coupled systems, one can expect a similar scenario 
that the stability of the vacuum also depends crucially on the presence of  the spontaneous annihilation process. 
Furthermore, as $N$ increases, the vacuum becomes more unstable in the $N$-BAW$_1$, because
the annihilation process of only the same species of particles is allowed. A sufficiently strong interspecies
coupling may wash away completely the absorbing phase for large enough $N$  even in one dimension, even
without any HC interaction. In this paper, we address this vacuum stability question in the strong coupling 
regime and measure the threshold value of $N$ with and without HC interaction. 

Outline of this paper is as follows. In Sec.~II, the $N$-species CP models are introduced.  
We find that, for any $N$, the vacuum is stable for a low branching rate 
and the DP-type phase transition is observed regardless of the presence of HC interaction. 
In Sec.~III, we study the $N$-species BAW$_1$ models by numerical simulations. We find that, for $N\ge 3$,
the vacuum becomes unstable with respect to an arbitrarily small branching
rate in a sufficiently strong coupling regime. The critical behavior near the zero
branching point is explored in terms of the scaling exponents. In a weak coupling regime, 
the conventional DP transition into a stable vacuum is observed. 
The interspecies HC interaction only shifts the DP critical points and the threshold value of
the interspecies coupling for complete disappearance of a stable vacuum. The $N=2$ case is special.
Without  HC interaction, one can always find the DP transition into a stable vacuum 
even at the full coupling strength, like in the $N$-species CP models. However, the inclusion of 
HC interaction makes the vacuum unstable again in a strong coupling regime.
We also discuss the effect of the directional asymmetry in the bidirectional couplings on
the phase diagram. Finally, we conclude in Sec.~IV.

\section{Coupled Contact Processes}

The $N$-species coupled contact process ($N$-CP) is defined  by the following evolution rules:
(1) Each particle annihilates spontaneously with probability $p$ 
or (2) creates a particle of the same species with probability $\sigma(1-p)$ or
a particle of the different species with probability $\sigma^\prime (1-p)/(N-1)$ 
in its neighborhood (branching process) as
\begin{equation}
\label{eq-1}
\begin{array}{llll}
A_i           & \longrightarrow & \emptyset     & \mbox{with $p$,} \\
A_i           & \longrightarrow & A_i + A_i     & \mbox{with $\sigma(1-p)$,} \\
A_i           & \longrightarrow & A_i + A_j     & \mbox{with $\sigma^\prime (1-p)/(N-1)$,}
\end{array}
\end{equation}
where $\sigma^\prime = 1-\sigma$, $i \neq j$, and $i=1,\ldots,N$.
Any branching attempt is rejected if it would result in a multiple occupation by the same
species  particles at a site. With interspecies HC interaction, a multiple 
occupation by different species particles is also forbidden. 
Each species is symmetrically coupled with other species 
by the interspecies coupling strength $\sigma^\prime$.
For $\sigma^\prime =0$, all species are completely decoupled.

It is trivial to show that the $N$-CP models with HC interaction become 
identical to the single-species CP model. Each site can be occupied by only one particle,
regardless of its species. Any configuration and any dynamic process in the N-CP
can be exactly mapped on its corresponding configuration and dynamic process of the
single-species CP by simply ignoring the particle species. 

\begin{figure}
\includegraphics[scale=0.4]{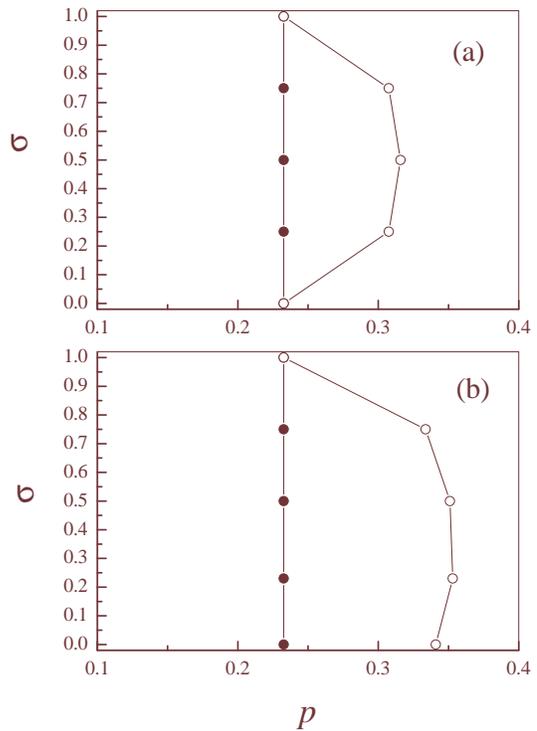}
\caption{\label{fig-1}
The $\sigma - p$ phase diagram for the $N$-CP, (a) $N=2$ and (b) $N=3$.
Filled and open circles correspond 
to critical points with and without HC interactions, respectively. Lines between data 
points are guides to the eyes only.}
\end{figure}

Without HC interaction, a multiple occupation by different species particles is allowed.
One can easily expect that the active phase expands until the multiple occupation is
maximized. For $N=2$, there exists a simple duality between two species 
$(\sigma \leftrightarrow \sigma^\prime)$ and the maximum point is located 
at $\sigma=\sigma^\prime=1/2$, see Fig.~\ref{fig-1} (a). 
For $N\ge3$, the maximum point is expected to be at $\sigma =1/N$,
where the branching symmetry is perfect and the maximum mixing is expected,
see Fig.~\ref{fig-1} (b). From this reasoning, we expect that complete
disappearance of the vacuum does not occur at any coupling strength $\sigma$ 
for any finite $N$. Of course, all absorbing critical phenomena should belong
to the DP class. 

\section{Coupled BAW$_1$}

The $N$-species coupled branching annihilating random walks with one offspring 
($N$-BAW$_1$), is defined by the following evolution rules:
(1) Each particle hops to a nearest-neighboring site with probability $p$ 
or (2) creates a particle of the same species with probability $\sigma(1-p)$ or
a particle of the different species with probability $\sigma^\prime (1-p)/(N-1)$ 
in its neighborhood (branching process) as

\begin{equation}
\label{eq-2}
\begin{array}{llll}
A_i \emptyset       & \longleftrightarrow & \emptyset  A_i     & \mbox{with $p$,} \\
A_i           & \longrightarrow & A_i + A_i     & \mbox{with $\sigma(1-p)$,} \\
A_i           & \longrightarrow & A_i + A_j     & \mbox{with $\sigma^\prime (1-p)/(N-1)$,}
\end{array}
\end{equation}
where $\sigma^\prime = 1-\sigma$ and  $i \neq j$.
If two identical particles happen to be on the 
same site, both particles immediately annihilate each other 
($A_i +A_i \longrightarrow \emptyset $).
With interspecies HC interaction, any hopping or branching attempt is rejected if
it would result in a multiple occupation of particles, regardless of their species.
The above model has a permutational symmetry between species. 
Later, we will also study an asymmetrically coupled system which breaks the permutational
symmetry.

\subsection{$N=2$}

\begin{figure}
\includegraphics[scale=0.4]{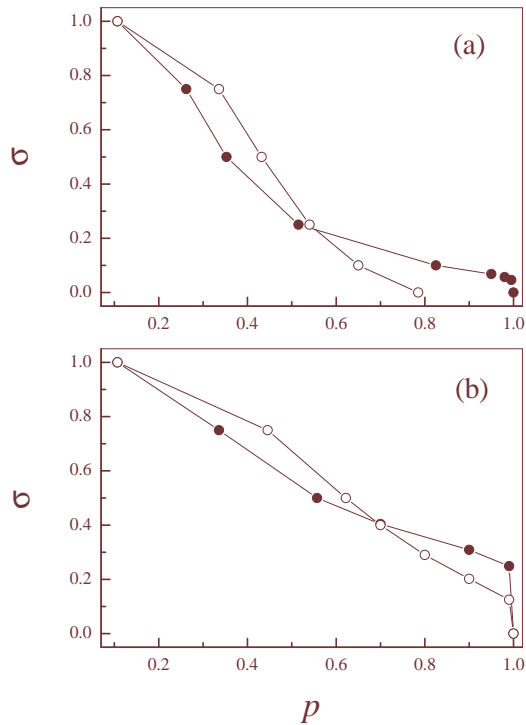}
\caption{\label{fig-2}
The $\sigma - p$ phase diagram of the $N$-BAW$_1$, (a) $N=2$ and (b) $N=3$. 
Filled and open circles correspond 
to critical points with and without HC interactions, respectively. Lines between data 
points are guides to the eyes only.}
\end{figure}

The two-species BAW$_1$ model has been investigated previously for general $\sigma$ 
\cite{KP1}. In this paper, we present the numerically improved phase diagram
in Fig.~\ref{fig-2} (a) and briefly summarize the results of \cite{KP1} for 
comparison to those for $N\ge 3$ and also for self-containedness.
Without HC interaction, the system always exhibits the DP-type absorbing transition for
all $\sigma$, including the case of the maximum coupling strength ($\sigma^\prime = 1$
or $\sigma=0$). 
However, with HC interaction, there exists a strong coupling regime 
($\sigma^\prime > \sigma^{\prime *}$) where the vacuum becomes unstable with respect to
an arbitrarily small branching rate ($p=1^{-}$) and completely disappears in the phase diagram
except at the annihilation fixed point ($p=1$). 
The threshold value for the strong coupling regime is numerically estimated
as $\sigma^{\prime *}\simeq 0.95$.  In the weak coupling regime ($\sigma^\prime < \sigma^{\prime *}$),
we observe the conventional DP transition into the vacuum.  

It is not surprising to see that the active phase expands as $\sigma^\prime$ 
increases, because a collision chance of the same species particles decreases.
However, the effect of HC interaction is rather tricky. For small $\sigma^\prime$,
the system tends to form large domains of the same species particles. The HC interaction
induces an effective diffusion bias directed to the domain center, which accelerates the pair 
annihilation process. Therefore the system becomes less active with HC interaction.
For large $\sigma^\prime$, the situation is reversed. The system prefers
locally heterogeneous configurations and the HC interaction reduces a chance of
binary collision of the same species particles. Thus, in this case, the system becomes
more active with HC interaction. This explains why two critical lines with and without
HC interaction should cross each other as in Fig.~\ref{fig-2}.

Consider the $\sigma = 0$ line, where a particle cannot create a particle of the same species.
As a result, a single particle cannot be annihilated by a single branching and diffusion process:
$A\rightarrow A + A  \rightarrow \emptyset $. It needs at least three branching processes such as
$A\rightarrow AB \rightarrow ABA \rightarrow ABAB$ or $ABBA$. This four particle state
can turn into vacuum via successive pair annihilations only if diffusion across a different
species particle is allowed (no HC interaction).  Hence, the vacuum can be stable in a low 
branching (high diffusion) regime. However, in the presence of HC interaction, the ordered
$AB$ pairs can not be annihilated by diffusions. Therefore, a single particle has a nonzero probability to
survive asymptotically and the vacuum is unstable with respect to an arbitrarily small
branching rate. 

The critical behavior near the annihilation fixed point in the strong coupling regime
is characterized by a set of the scaling exponents
\begin{equation}
\label{eq-3}
\beta = 1/2,\; \nu_{\bot} = 1/2,\; \nu_{||} = 1, 
\end{equation}
where the exponents $\beta$, $\nu_\bot$, and $\nu_{||}$
characterize the scaling behavior of the steady-state particle density $\rho_s$, 
the correlation length $\xi$, and the relaxation time $\tau$, respectively.

The exponent ratios, $\beta/\nu_\bot=1$ and $z=\nu_{||}/\nu_\bot =2$, originate from
the ordinary diffusion nature at the annihilation fixed point in one dimension \cite{KLP}. 
There is only one independent exponent $\beta$, of which the value can be extracted 
by a simple argument \cite{KP1}. Consider a particle $A$ created by a particle $B$.
This branching process  increases the particle density with the time scale
$\tau_b \sim (1-p)^{-1}$. Near the annihilation fixed point of zero branching rate,
the offspring $A$ would be annihilated by colliding with an independent $A$ via diffusion.
The time scale for this process is governed by ordinary diffusion:
$\tau_d \sim \ell^2$ where the mean distance between particles $\ell$ is order of 
the inverse of the particle density $\rho^{-1}$. 
Balancing these two time scales, we can expect the steady-state particle density to scale as
$\rho_s \sim (1-p)^{\beta}$ with $\beta = 1/2$.

This argument is quite general, so it should apply to many other models exhibiting
a critical behavior near the annihilation fixed point.  Moreover, the HC interaction
does not matter in this argument. So, one can expect that the $N$-species BAW$_1$ should
belong to the same class ($\beta = 1/2$) for any $N\ge 2$, regardless of the presence of
HC interaction, which will be confirmed numerially in next subsection. Moreover,  
the $N$-species BAW$_2$ also belongs to the same class when the HC interaction is
present and the branching method is {\em static}: Two offspring are divided by their parent, 
$\emptyset B \emptyset \rightarrow ABA $. It is clear that our argument applies to this model,
due to the HC interaction. When the branching method is {\em dynamic} or there is no HC
interaction, $N$-BAW$_2$ belongs to different universality classes \cite{KLP}.

\subsection{$N\ge 3$}

First, consider the $N=3$ case without HC interaction. 
To map out the phase diagram, we perform the defect-type dynamic Monte 
Carlo simulations, starting with a single particle.
We measure the survival probability  $P(t)$,
the number of particles $N(t)$, the mean distance of spreading $R(t)$. 
At criticality, these quantities scale algebraically in the long time limit 
as $P(t) \sim t^{-\delta}$, $N(t) \sim t^{\eta}$ and $R(t) \sim t^{1/z}$ \cite{Grass}. 

By inspecting the curvature of effective exponents defined as
\begin{equation}
\label{eq-4}
-\; \delta(t) = \log[P(t)/P(t/m)]/\log m ,
\end{equation}
with arbitrary $m$, and similarly for $\eta(t)$ and $1/z(t)$, 
we estimate the values of the critical hopping probability $p_c$ and
the dynamic exponents $\delta$, $\eta$ and $z$ for various values of $\sigma$.
The $\sigma-p$ phase diagram is shown in Fig.~\ref{fig-2} (b).
For small $\sigma$, the absorbing phase (vacuum) completely disappear.
This result is rather unexpected because, for $N=2$,  the vacuum is always stable 
for a low branching rate without HC interaction. Before going into detailed discussion on 
this vacuum instability without HC interaction, we present numerical results.
The threshold value of the vacuum instability is estimated as
$\sigma^{*} \simeq 0.125$ $(\sigma^{\prime *}\simeq 0.875)$.
For $\sigma > \sigma^* $, we observe that the system undergoes 
an absorbing transition into vacuum. As expected, we find the DP critical
exponents along both paths of constant $p$ and constant $\sigma$ lines.

To identify the scaling behavior near the annihilation fixed point $(p=1)$ for 
$\sigma < \sigma^* $,  we analyze the finite-size
effects on the steady-state particle density $\rho_s$.
Using the finite-size scaling theory on $\rho_s$ \cite{Aukrust}
\begin{equation}
\label{eq-5}
\rho_s (\Delta,L) = L^{-\beta/\nu_{\bot}} F(\Delta L^{1/\nu_{\bot}} )
\end{equation}
with  $\Delta=p_c - p$ and system size $L$, 
the value of $\nu_{\bot}$ is determined by collapsing the data of $\rho_s$ with 
$\beta/\nu_{\bot} = 1$.
We measure $\rho_s$ in the steady state, averaged over $5 \times 10^3 \sim 5 \times 10^4 $ samples
for several values of $\Delta$ ($5 \times 10^{-4} \sim 0.05$) and $L$ ($2^5 \sim 2^9 $).

\begin{figure}
\includegraphics[scale=0.6]{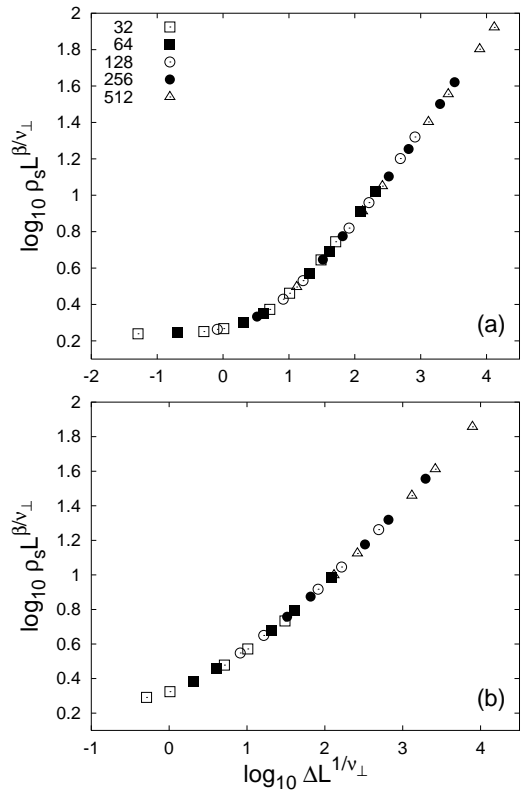}
\caption{\label{fig-3}
Data collapse of $\rho_s L^{\beta/\nu_{\bot}}$ against $\Delta L^{1/\nu_{\bot}}$ with
$\beta/\nu_{\bot}=1$ and $\nu_\bot=1/2$ for system size $L=2^5, \cdots, 2^9$ for $N=3$ (a) without
HC interaction at $\sigma = 0$ (b) with HC interaction at $\sigma =0.2$. 
}
\end{figure}

We estimate $\nu_{\bot} = 0.50(4)$ at $\sigma = 0$. 
Fig.\ref{fig-3} (a) shows very good data collapse of $\rho_s$ with $\nu_{\bot} = 0.50$.
These numerical results confirm our previous argument for the universality class near
the annihilation fixed point. 

To understand the vacuum instability for small $\sigma$ for $N=3$ without HC interaction, 
we again consider the $\sigma=0$ line, where a particle can create a particle of different
species only. As explained before, there needs a sequence of branching processes 
for a single particle annihilation. For $N=2$ without HC interaction, the process of
$A\rightarrow AB \rightarrow ABA \rightarrow ABAB\rightarrow AA \rightarrow \emptyset$
is possible. In this case, $AA$ or $BB$ pairs will quickly go away at a very high diffusion rate,
so one may consider only $AB$ pairs, which cannot be annihilated by itself.
The last two reaction events can be regarded as a collision and annihilation event of two $AB$ pairs
and the previous reaction events as a creation event of a $AB$ pair out of a parent $AB$ pair.
This process is exactly equivalent to the $N=1$ BAW$_1$ model by identifying a $AB$ pair
as a particle. Therefore the $N=2$ case without HC interaction can have a stable vacuum like
in the $N=1$ model. The $N=3$ case is different. Three different pairs ($AB$, $BC$,  $CA$ )
are possible. Any pair can branch any other pair,  
e.g.~$AB \rightarrow ACBA \rightarrow CB\rightarrow CBBA \rightarrow CA$. 
So, the $N=3$ case can not reduce to the $N=1$  model, in contrast to the $N=2$ case.
It seems that this multispecies character is the crucial element for vacuum instability
in BAW-type binary diffusion-annihilation models. 
 
Interspecies HC interaction should destabilize the vacuum more easily. We expect that 
the threshold of vacuum instability should be lowered with HC interaction.
In Fig.~\ref{fig-2} (b), the phase diagram for $N=3$ with HC interaction is presented. 
We estimate $\sigma^{*} \simeq 0.25$. 
To estimate $\nu_{\bot}$ for $\sigma < \sigma^* $, 
we also try the data collapse of $\rho_s $.  In Fig.~\ref{fig-3} (b), 
we present the data at $\sigma=0.2$ with the exponent values of
$\nu_\bot = 0.50$ and $\beta/\nu_\bot =1$.  Our estimation
is $\nu_\bot = 0.50(2)$. Again, we find that our simple argument 
also applies to this case. 

We also perform defect-type Monte Carlo simulations for the $N=4$ case. 
We estimate $\sigma^{*} = 0.22(2)$ without HC interaction and $\sigma^{*} = 0.328(2)$ 
with HC interaction. The exponent value is estimated as 
$\nu_\bot\simeq 0.52(2)$ at $\sigma=0.15$ without HC interaction and
$\nu_\bot\simeq 0.52(2)$ at $\sigma=0.3$ with HC interaction.

\subsection{asymmetric coupling}

We study the effect of broken permutational symmetry on the phase diagram.
For convenience, we consider the $N=3$ case at $\sigma=0$ ($\sigma^\prime =1$) 
without HC interaction only. The branching process in Eq.~(\ref{eq-2}) is modified as
\begin{equation}
\label{eq-6}
\begin{array}{lllllll}
A        & \longrightarrow & A + B , &B        & \longrightarrow & B + A &\mbox{with $(1-q)(1-p)$,} \\
A        & \longrightarrow & A + C , &B        & \longrightarrow & B + C &\mbox{with $q(1-p)$,} \\
C        & \longrightarrow & C + A , &C        & \longrightarrow & C + B &\mbox{with $(1-p)/2$,} \\
\end{array}
\end{equation}
where $0\le q \le 1$. 

At $q=0$, the species $C$ is completely suppressed and the model becomes identical to
the $N=2$ symmetric one at $\sigma=0$. At $q=1/2$, all three species are equivalent
and the $N=3$ symmetric model is recovered. Our results on the symmetric $N$-BAW$_1$ models
in previous subsections indicate that the vacuum is stable in  high diffusion regime
at $q=0$, but becomes unstable completely at $q=1/2$. For $q<1/2$. the species $C$ is
suppressed in comparison to the other two species $A$ and $B$. Therefore,
the density of the third species $C$ should be proportional to $q$. Here, we try to locate the 
threshold value of $q$ for complete vacuum instability. 

\begin{figure}
\includegraphics[scale=0.6]{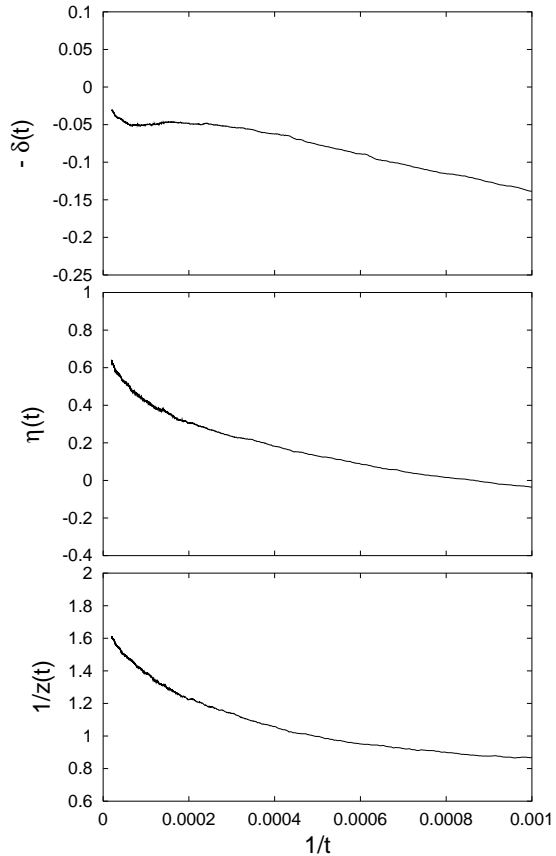}
\caption{\label{fig-4}
Plots of the effective exponents against $1/t$ at $p=0.999$ for $q=0.1$.
Upward curvature of each exponent indicates that the system is still in the active
phase at $p=0.999$.
}
\end{figure}

We perform  the defect-type  dynamic simulations for several $q$ values ($ 10^{-1} \sim 1.0$) to locate $p_c$. 
In Fig.~\ref{fig-4}, we present simulation results for $q=0.1$ at $p=0.999$.
Effective exponents $\delta (t)$, $\eta (t)$ and $1/z(t)$ show upward curvatures, which  
imply that  the system is still active even at $p=0.999$. 
It suggests that the criticality is located at $p=1.000(1)$ and the vacuum is completely unstable.
For other nonzero $q $ values, we also find the similar results to those for $q=0.1$. 
We conclude that, for any $q \neq 0$, the system is always active and only critical at $p_c = 1.0$. 
As discussed in the previous subsection, this result again confirms that the multispecies character 
is relevant (not the symmetry) to vacuum instability. We also check the $C$-dominant regime for $q>1/2$
and find a similar result. 

\begin{figure}
\includegraphics[scale=0.6]{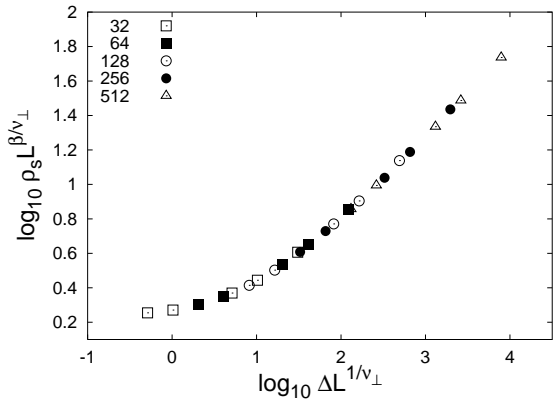}
\caption{\label{fig-5}
Data collapse of $\rho_s L^{\beta/\nu_{\bot}}$ against $\Delta L^{1/\nu_{\bot}}$ with
$\beta/\nu_{\bot}=1$ and $\nu_\bot=1/2$ for $q=0.1$. 
The steady state density $\rho_s $ is measured for various system size $L$ ($2^5 \sim 2^9$)
and $\Delta$ ($5 \times 10^{-4} \sim 3 \times 10^{-2}$).
}
\end{figure}

To identify the critical behavior, we estimate the exponent $\nu_\bot$ by 
collapsing the particle density data with $\beta/\nu_\bot =1$.
Fig.~\ref{fig-5} shows the scaling plot for $q=0.1$ with $\nu_\bot = 0.5$. 
We estimate $\nu_\bot = 0.50(5)$ for $q=0.1$. 
For other nonzero $q$ values, we also estimate $\nu_\bot \simeq 0.5$.

Although we do not consider the incomplete coupling case ($\sigma \neq 0$), we
expect a nonzero threshold value of $\sigma^*$, below which the system is always active.

\section{conclusion}

In this paper , we study the stability of vacuum in $N$ coupled DP systems.
The interspecies coupling is linear, bidirectional, and excitatory.
In the coupled contact processes, the vacuum is always stable at a sufficiently low branching rate
for all $N$, regardless of the coupling strength $\sigma^\prime$, and the system undergoes DP type absorbing 
transition into the vacuum. On the other hand, in the coupled BAW with one offspring, the vacuum
stability is quite fragile for $N\ge 2$ in a strong coupling regime. The absence of a spontaneous
annihilation process $A\rightarrow\emptyset$ is crucial for vacuum instability.

We find that the vacuum is unstable with respect to an arbitrarily small branching rate 
in a sufficiently strong coupling regime ($\sigma\prime > \sigma^{\prime *}$) for $N\ge 3$. 
The multispecies character is the key element responsible for this vacuum instability 
and the asymmetry in the interspecies coupling is shown to be irrelevant. 
The $N=2$ case is special. The vacuum is always stable like in the coupled contact processes,
but the HC interaction is relevant to vacuum instability in a strong coupling regime.
We show that the $N=2$ BAW$_1$ model without HC interaction can reduce to the $N=1$ model even at the
full coupling strength, which explains the specialty at $N=2$.

Critical behavior near the annihilation fixed point  in a strong coupling regime can be conjectured
by a simple argument of balancing two time scales of branching and annihilating random walks. 
Numerical investigations confirm our conjecture of $\beta=1/2$, $\nu_\bot =1/2$, and 
$\nu_{||}=1$, which also applies to $N$-BAW$_2$ with static branching and HC interaction \cite{KLP}.


\end{document}